\documentclass[12pt,preprint]{aastex}

\begin{document}
\newcommand{\pul}{PSR~J0538+2817}
\newcommand{\xmm}{{\it XMM-Newton}}
\newcommand{\chandra}{{\it Chandra}}

\shorttitle{Pulsed X-ray Emission from {\it XMM-Newton}}
\shortauthors{McGowan et al.}

\title{Detection of Pulsed X-ray Emission from {\it XMM-Newton}
Observations of PSR~J0538+2817}
\author{K.E. McGowan\altaffilmark{1}, J.A. Kennea\altaffilmark{2},
S. Zane\altaffilmark{3}, F.A. C\'{o}rdova\altaffilmark{4}, 
M. Cropper\altaffilmark{3}, C. Ho\altaffilmark{1}, \\ 
T. Sasseen\altaffilmark{2}, W.T. Vestrand\altaffilmark{1}}
\altaffiltext{1}{Los Alamos National Laboratory, MS D436, Los Alamos, NM 87545}
\altaffiltext{2}{University of California, Santa Barbara, CA 93106}
\altaffiltext{3}{Mullard Space Science Laboratory, University College of
London, UK}
\altaffiltext{4}{University of California, Riverside, CA 92521}
\email{mcgowan@lanl.gov}

\begin{abstract}
We report on the \xmm\ observations of the 143~ms pulsar \pul.  We
present evidence for the first detections of pulsed X-rays from the
source at a frequency which is consistent with the predicted radio
frequency.  The pulse profile is broad and asymmetric, with a pulse
fraction of $18 \pm 3$\%.  We find that the spectrum of the source is 
well-fit with a blackbody with $T^{\infty} = (2.12^{+0.04}_{-0.03})
\times 10^6$ K and $N_{H}$ = 2.5$\times 10^{21}$ cm$^{-2}$.  The
radius determined from the model fit of $1.68\pm0.05$~km suggests that
the emission is from a heated polar cap.  A fit to the spectrum with an 
atmospheric model reduces the inferred temperature and hence increases
the radius of the emitting region, however the pulsar distance
determined from the fit is then smaller than the dispersion distance. 
\end{abstract}


\section{INTRODUCTION}
\renewcommand{\thefootnote}{\fnsymbol{footnote}}
\setcounter{footnote}{0}

Of the $> 1000$ radio pulsars detected so far, 51 have also been found in
the X-rays \cite{beck02a}.  These X-ray emitting pulsars represent a
wide range of ages ($10^3-7\times 10^9$~yrs), magnetic field strengths 
($10^8-10^{13}$~G), periods (1.6 -- 530~ms) and spectral properties.
In particular, while non-thermal (power law) emission is supposed to
dominate in the youngest objects (like the Crab), the soft X-ray
radiation of rotation-powered pulsars with ages as large as 
$\approx 10^5-10^6$~yrs should be dominated by thermal emission from
the neutron star (NS) surface (see e.g. Becker \& Pavlov 2002 for a 
recent review).  These ``middle-aged pulsars'' are old enough for
their magnetospheric emission to have decreased and become fainter
than the thermal one.  At the same time, they are young enough that
the NS surface is not too cool for its thermal radiation to still be 
detectable in the X-rays.

There are three middle-aged pulsars for which a thermal component,
interpreted as thermal emission from the surface, has certainly been
observed: Geminga, PSR~B0656+14 and B1055-52 (dubbed the {\it three
Musketeers} by Becker \& Tr\"{u}mper 1997).  Due to its age, a
dominant thermal component was also expected from \pul.

\pul\ is a 143~ms pulsar located in the direction of the Galactic anti-center
\cite{and94}.  The source was detected during an untargeted pulsar
survey at 430~MHz using the 305~m Arecibo radio telescope
\cite{fos95}.  \pul\ is located at a distance of 1.2~kpc (using the
model of Cordes \& Lazio 2002), with a spin down age of
$6\times10^{5}$~yrs.  The source is believed to be associated with the
supernova remnant S147 \cite{and96} as the distance to the pulsar and
the remnant are consistent, as are their ages.  Sun et al.\ (1996)
detected X-ray emission from \pul\ using data from the ROSAT All Sky
Survey (RASS).  However, the photon statistics from the RASS were not
good enough to perform timing or spectral studies.  Marginal evidence
was found for pulsed gamma-ray emission at the radio period from
photon arrival time analysis of EGRET data \cite{sun96}.

\pul\ is a factor 5--10 more distant than Geminga, and only the large
collective area of \xmm\ has now made it possible to obtain good spectral
information.  In this paper we report on \xmm\ observations of \pul\
and present timing and spectral analyses of the source.

\section{OBSERVATIONS}

\pul\ was observed with \xmm\ for $\sim$18~ks on 2002 March 8 as a part
of the Guaranteed Time Program.  The following analysis uses data from 
the three European Photon Imaging Camera (EPIC) instruments: two EPIC
MOS detectors \cite{tur01} and the EPIC PN detector \cite{stru01}.  
The thin optical blocking filter was used on the PN.  To provide a
temporal resolution of 6~ms, which is needed for timing studies of
millisecond pulsars, the PN was operated in {\it small window} mode.
The MOS1 was operated in {\it full window} (imaging) mode with a time
resolution of 1.4~s.  In order to obtain better temporal resolution
the MOS2 was operated in {\it timing mode}; in this mode data from the
central CCD are collapsed into a one-dimensional row to achieve a
1.5~ms time resolution.  The medium filter was used for both MOS
observations. 

Observations of \pul\ were taken with the Resolution Grating
Spectrometer (RGS), however we did not detect enough photons for a
meaningful analysis.  We also do not detect the pulsar in the images
taken with the OM.  We are unable to determine a realistic upper limit
for the optical detection of \pul\ as the position of the pulsar is
coincident with the stray light effect in the OM image.

We reduced the EPIC data with the \xmm\ Science Analysis System (SAS
v5.3.0).  To maximize the signal-to-noise we filtered the data to
include only single, double, triple and quadruple photon events for
MOS, and only single and double photon events for PN.  We find no
evidence for any extended emission in a spatial analysis of the MOS
and PN data, indicating that \pul\ is consistent with being a point
source from \xmm\ observations.  

We extracted data for the source within a radius of 30$\arcsec$ for
the PN and MOS1 instruments.  While this radius encircles 100\% of the
flux for \pul\ in the MOS1, only 89\% of the flux is encompassed for
the PN.  However, due to the position of the source on the PN detector
we are restricted in our choice of radii.  Hence, the flux measured
from the PN is corrected for in the following analyses.  Selection of
the photons from the MOS2 event file was achieved by extracting 100\%
of the flux within a rectangular region centered on the source. 

In our temporal analysis we consider data within the 0.2--15~keV
energy range for the EPIC MOS2 and PN instruments.  The timing studies
were performed on barycentrically corrected MOS2 and PN event files. 

We used MOS1 and PN data to produce spectra of \pul.  We subtracted a 
background which was extracted from an annulus around the source
fiducial region.  We also created the corresponding photon
redistribution matrix (RMF) and ancillary region file (ARF). 

\section{TIMING ANALYSIS}

We determine a predicted pulse period for \pul\ from the radio
measurements \cite{and96} at the epoch of our \xmm\ observations (MJD
2452341.5), assuming a linear spin-down rate.  It is possible however that
glitches could alter the period from the predicted value.  Therefore
we searched for a pulsed signal in the PN ($\sim$6500 photons) and 
MOS2 ($\sim$3300 photons) data over a wider frequency range centered on
the derived radio period, P = 143.158657~ms ($f=6.9852569$ Hz).  

We have employed two methods in our search for pulsed emission from
\pul.  In the first method we implement the $Z^{2}_{n}$ test
\cite{bucc83}.  In the second method we calculate the Rayleigh
Statistic (see de Jager 1991; Mardia 1972) and then calculate the
Maximum Likelihood Periodogram (MLP) using the $\Delta C$-statistic
\cite{cash79} to determine significant periodicities in the datasets
(see Zane et al.\ 2002).  The advantage of this method is that the
$\Delta C$-statistic makes no assumptions about the distribution of
the data, and is itself distributed as $\chi^{2}$ which allows one to
determine errors for the periods detected directly from the periodogram.  

\begin{figure}[h]
\vbox to5in{\rule{0pt}{5in}}
\includegraphics{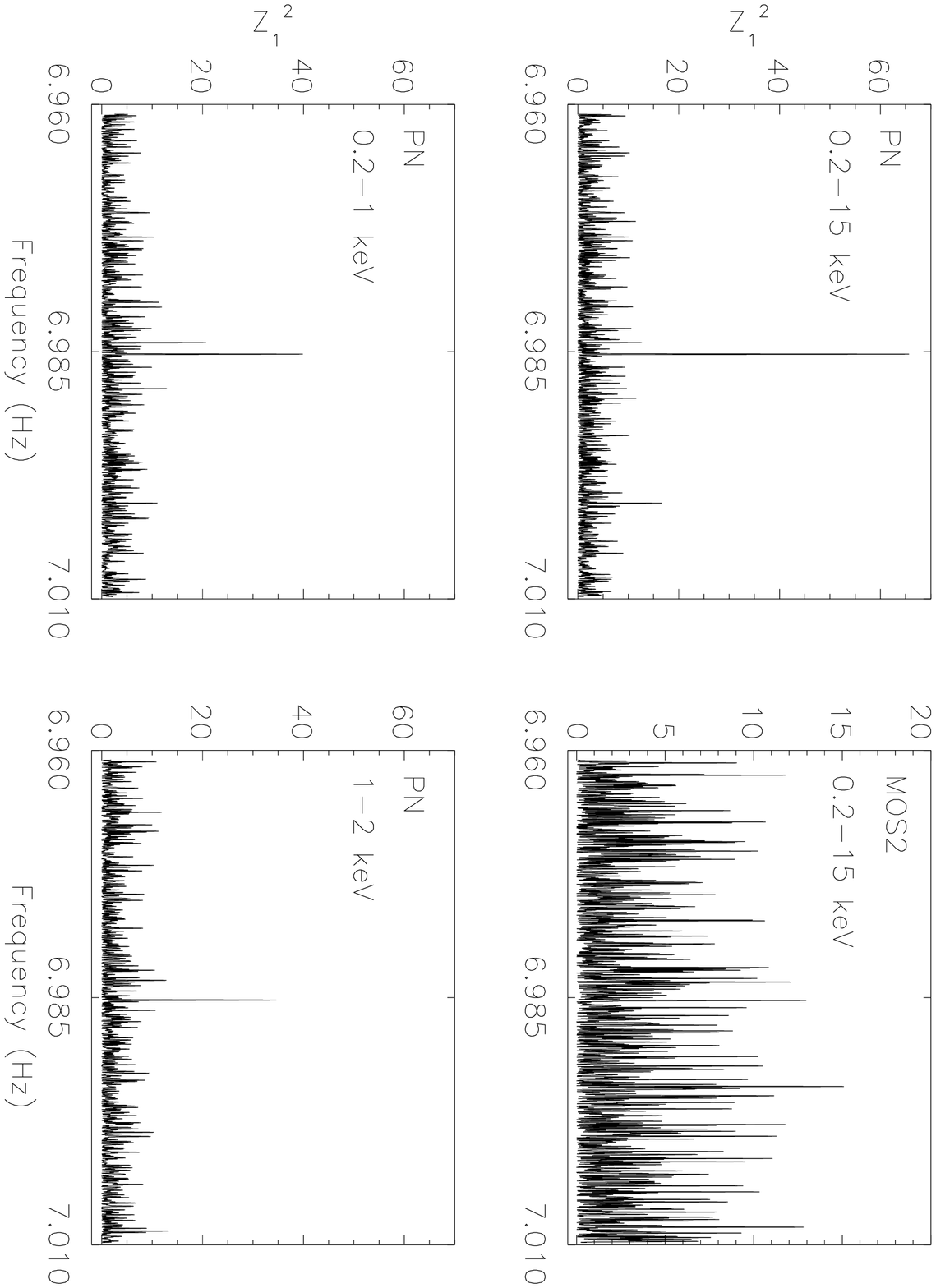}
\caption{$Z^{2}_{1}$ periodograms for \pul.  Top, left panel, PN data
in the 0.2--15 keV energy range, top, right panel, MOS2 data in the
0.2--15 keV energy range.  Bottom, left panel, PN data in the 0.2--1
keV energy range, bottom, right panel, PN data in the 1--2 keV energy
range.}\label{fig:ztest}
\end{figure}

We searched the PN data (0.2--15 keV) for a periodic signal using the
$Z^{2}_{n}$ test, with the number of harmonics $n$ being varied from 1 to
10.  We find that in order to optimize the signal-to-noise ($S$/$N$)
ratio we must use the value $n=1$.  The resulting $Z^{2}_{1}$
periodogram is shown in Fig.~\ref{fig:ztest} (top, left panel).  We
find a peak at $f=6.9852548^{+0.0000104}_{-0.0000124}$ Hz, which,
within the quoted 90\% uncertainty range, is consistent with the
predicted radio frequency given above, $f = 6.9852569$ Hz.  The
$Z^{2}_{1}$-statistic for this peak is 65.67, which has a probability
of chance occurrence of $5.5 \times 10^{-15}$.  Frequencies of this
order are detected at the values expected, given radio measurements,
in similar objects (J.A. Kennea 2003, private communication).

The corresponding $Z^{2}_{1}$ periodogram for the MOS2 data (0.2--15
keV) is shown in Fig.~\ref{fig:ztest} (top, right panel).  We note
that the noise level is much greater in the MOS2 periodogram compared
to that for the PN data. The largest peak near to the predicted radio
frequency, and the second largest peak in the periodogram, is at
$f=6.9852939^{+0.0000117}_{-0.0000148}$ Hz.  This value, within the 90\%
confidence limit quoted, is not consistent with the frequency found in
the PN data, or the derived radio frequency.  If we consider the 68\%
uncertainty range, $f=6.9852939^{+0.0000243}_{-0.0000262}$ Hz, the
frequency is still not consistent with the predicted and PN
frequencies.  The probability of chance occurrence for this peak is 
$1.5 \times 10^{-3}$.  The largest peak in the MOS2 periodogram occurs at
$f=6.9939901^{+0.0000103}_{-0.0000133}$ Hz.  However, it is difficult to
reconcile this value with the predicted radio frequency.  Both peaks
have a low significance when compared with the noise level.

Assuming that the pulsed signal we have detected in the PN data is
real we have investigated the energy range over which we can detect a
periodic signal for the PN data.  We separated the PN data into 4 energy
bands, 0.2--1, 1--2, 2--3 and 3--10 keV, and generated $Z^{2}_{1}$ 
periodograms for each dataset.  Our results show that we can detect a
pulsed signal in the 0.2--1 and 1--2 keV ranges, but at energies
greater than 2 keV we do not find a dominant peak in the periodograms
near to the predicted radio frequency.  We find a peak at
$f=6.9852548^{+0.0000095}_{-0.0000139}$ Hz for the 0.2--1 keV data
(Fig.~\ref{fig:ztest}, bottom, left panel), and at
$f=6.9852499^{+0.0000127}_{-0.0000126}$ Hz for the 1--2 keV data
(Fig.~\ref{fig:ztest}, bottom, right panel).  Both of these
frequencies are consistent with the frequency found previously for the
PN data (0.2--15 keV), and the derived radio frequency, within the 90\%
uncertainty range quoted.  The probability of chance occurrence for the
0.2--1 and 1--2 keV peaks are $2.2 \times 10^{-9}$ and  $3.0 \times
10^{-8}$, respectively.  The frequencies being consistent within
different energy ranges is expected.  The lower amplitude and
significance of detection of pulsed emission in the separate energy
ranges is simply due to the fact that fewer events are detected.   
 
\begin{figure}[h]
\epsscale{1.1}
\plottwo{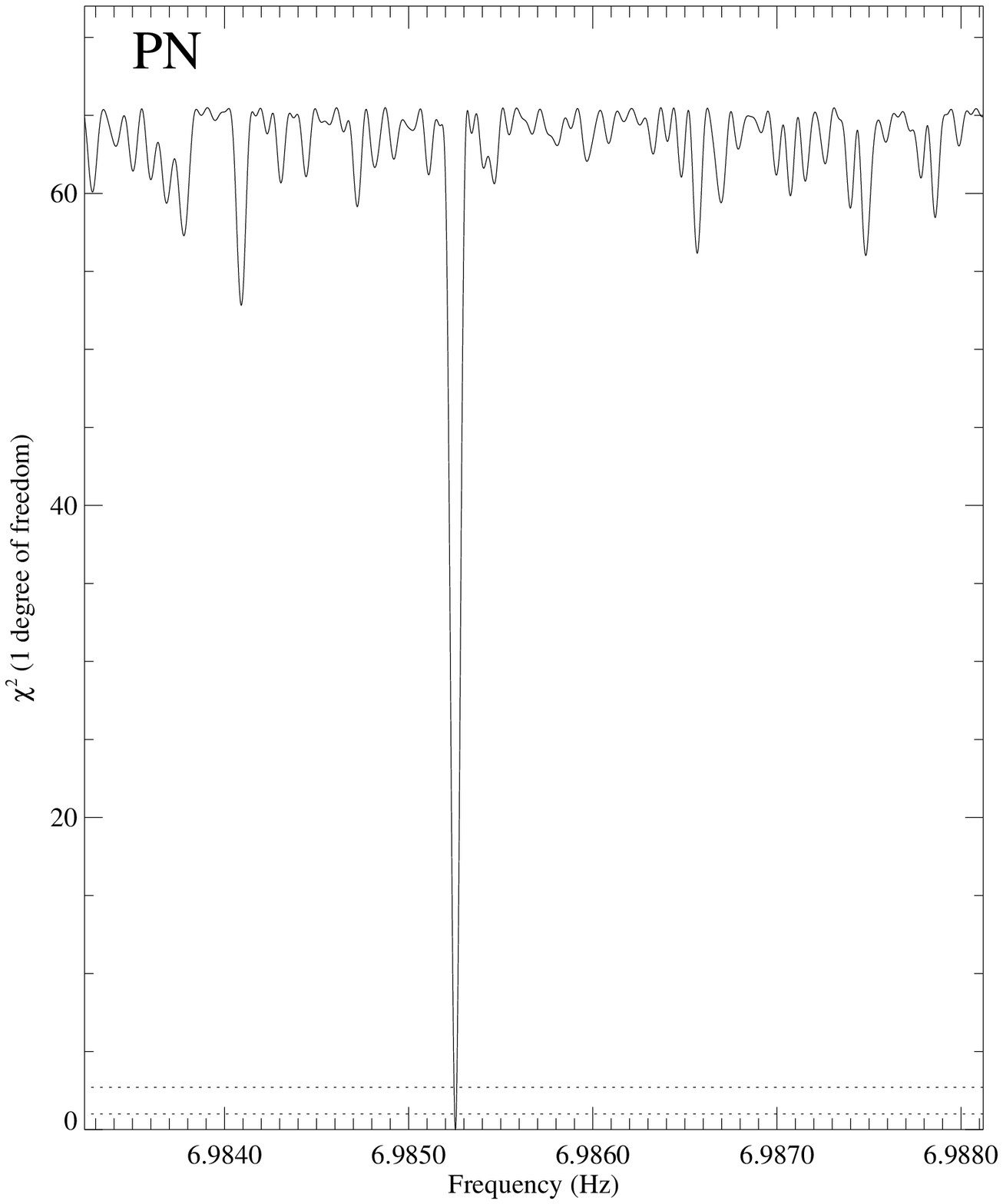}{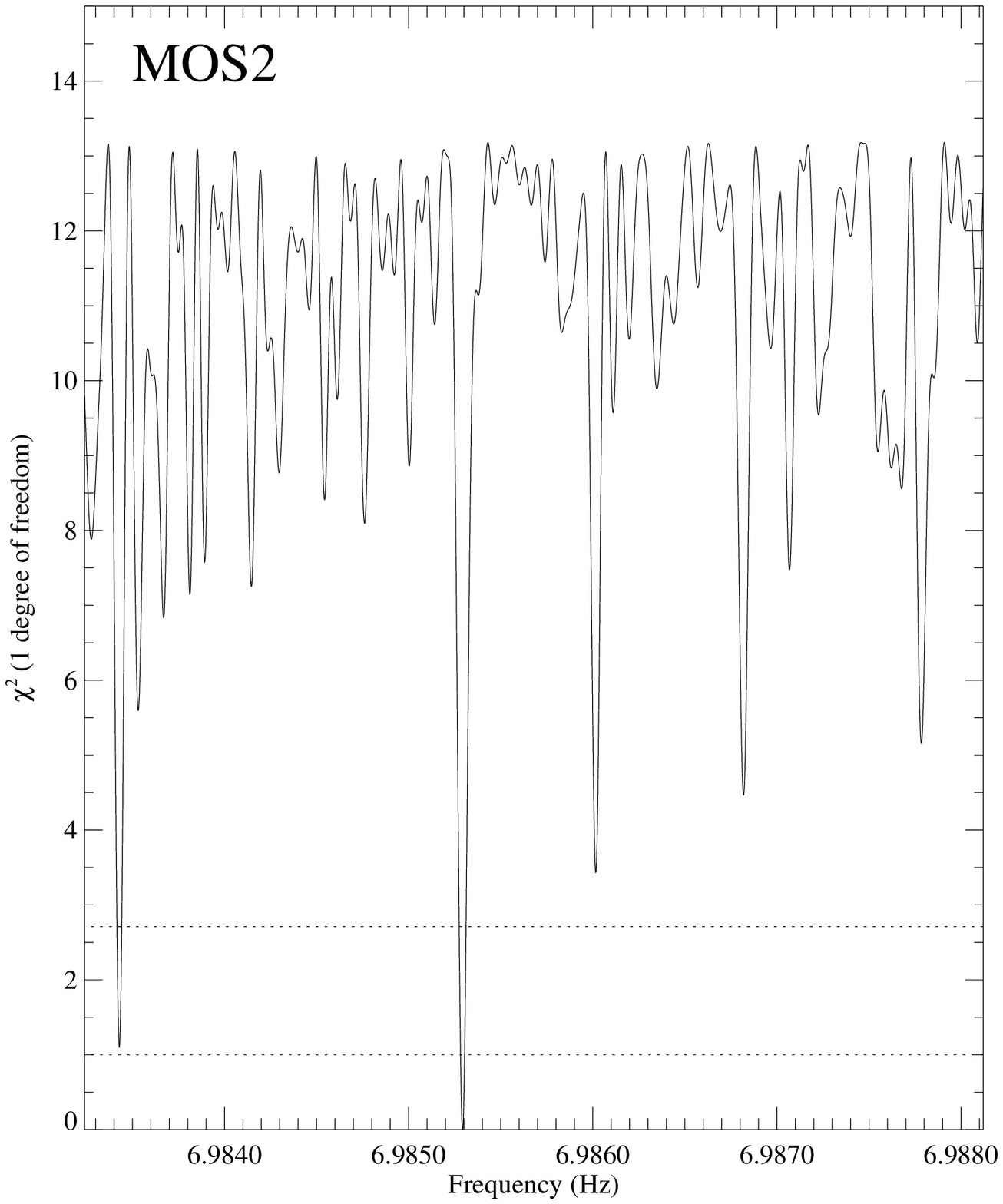}
\caption{Maximum likelihood periodograms (MLP) for the PN data (left) and
MOS2 data (right).  The peak in the MLP for the PN data is at
6.9852540~Hz.  In the MLP for the MOS2 data the peak nearest to the 
predicted frequency occurs at 6.9852931~Hz.  The 68\% and 90\% confidence
levels for the periods are at $\chi^2$ = 1.0 and 2.71 (dotted lines).}\label{fig:mlp}
\end{figure}

We also searched for a periodic signal using the MLP method.
Initially we constructed the MLPs using the PN and MOS2 data in the
0.2--15 keV energy range to compare with our $Z^{2}_{1}$ results.  We
find a peak for the PN data at $f=6.9852540^{+0.0000049}_{-0.0000039}$
Hz, which is detected at the 8-sigma level (Fig.~\ref{fig:mlp}, left
panel).  The frequency of the peak is consistent, within the 90\%
confidence limit, with the $Z^{2}_{1}$ value for the PN data, and with
the predicted radio frequency.

The largest peak in the MOS2 MLP occurs at
$f=6.9852931^{+0.0000112}_{-0.0000185}$ Hz (Fig.~\ref{fig:mlp}, right
panel).  While this is not consistent with the PN and predicted radio
frequencies, it is in agreement with the values found from the
$Z^{2}_{1}$ test for the MOS2 data.  While the peak could represent a
true periodicity, its significance above the noise is low. 

We used the MLP method on the PN data which had been separated into
different energy ranges, given above.  The results are in excellent
agreement with the $Z^{2}_{1}$ test, again with the significance and
amplitude of the detection decreasing with number of events.

While our results indicate that we have detected pulsed emission from
\pul\ in the PN data, the absence of a periodic signal at a similar
frequency, within the 90\% confidence limit, in the MOS2 data is
worrying.  However, we note that the MOS2 receives $\sim 40$\% of the
flux that is received by the PN, and that the quantum efficiency of
the MOS instruments is poorer than that for the PN, especially in the
softer energies where we are detecting the pulsed signal.  Despite
these factors, the most important consideration for this analysis is
the $S$/$N$ ratio for the PN and MOS2 data.  Our PN data has
$S$/$N\sim6$, while for the MOS2 we have $S$/$N\sim2.5$.  This
suggests that the lack of signal in the MOS2 data is hampering the
period search; our results for the MOS2 temporal analysis being
consistent with noise is perhaps not surprising.

The frequency found from the $Z^{2}_{1}$ test for the PN data in
the 0.2--15 and the 0.2--1 keV band, $f=6.9852548$ Hz, deviates from
the derived radio frequency by 0.0000021~Hz, or 0.000043~ms.  We have
folded the PN light curves in the 0.2--15, 0.2--1, 1--2 and 2--10~keV
bands using our detected frequency, and the radio ephemeris (MJD
2449564.5).  

\begin{figure}[h]
\epsscale{0.8}
\plotone{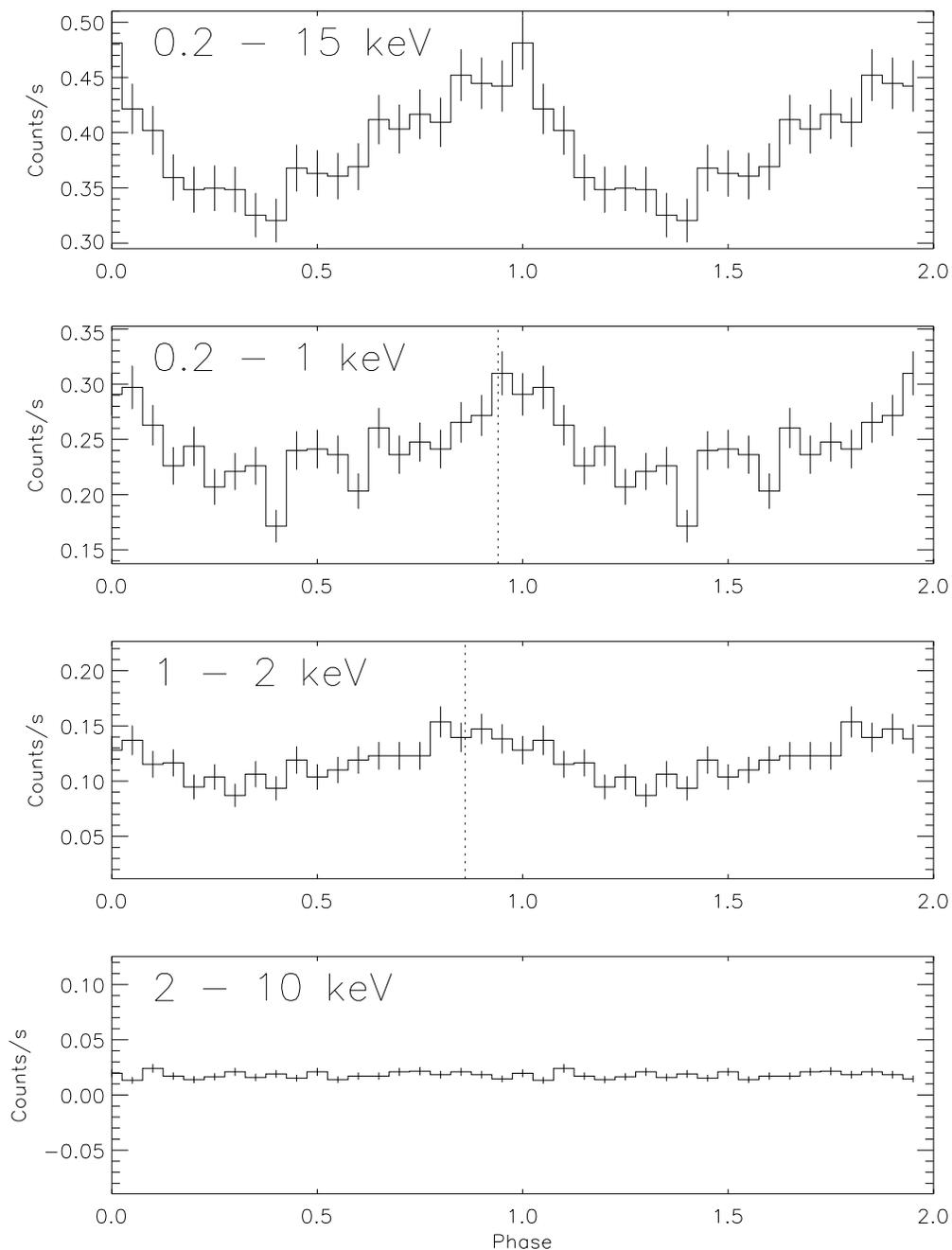}
\caption{Light curve of \pul\ from \xmm\ EPIC PN folded on $f=6.9852548$ Hz.
Light curves in the 0.2--15 (first panel), 0.2--1 (second panel), 1--2 
(third panel) and 2--10~keV bands (fourth panel).  The dotted lines in
the second and third panels show the position of the maximum of the
best fit sinusoid to each dataset, and correspond to $\phi=0.94$ and 
$\phi=0.86$, respectively.  These values indicate a shift in phase of 0.08.}\label{fig:lc}
\end{figure}

The 0.2--15~keV light curve (Fig.~\ref{fig:lc}, top panel) has a pulse 
fraction of $18\pm3$\%.  The pulse profile is clearly asymmetric with
a slower rise to and faster fall from maximum.  Indicative of this
asymmetry, a sinusoidal fit to the light curve can be excluded with
$\chi^{2}_{\nu} = 4.0$.  The arrival time of the radio pulse is
consistent with the maximum of the X-ray peak.  The light curves in
the 0.2--1 and 1--2~keV bands (Fig.~\ref{fig:lc}, second and third
panels) have pulse fractions of $30\pm3$\% and $27\pm3$\%,
respectively.   

In order to investigate the asymmetry in the 0.2--15~keV light curve
we determined whether or not there is a phase shift between the 0.2--1
and 1--2~keV light curves.  Although the PN light curve in the
0.2--15~keV energy range was not well described by a sinusoid, as a
first estimate to the maxima of the 0.2--1 and 1--2~keV light curves
we have employed the method of fitting with a sinusoid, as these two
light curves are more symmetrical.  While the fit for the 1--2~keV
light curve is improved by a factor of $\sim$ two the fit is still not
optimal, and the value of $\chi^{2}_{\nu}=3.6$ for the 0.2--1~keV
light curve still indicates an asymmetry.  However, for estimation
purposes we have marked the position of the maxima of the cosine
curves for each band in Fig.~\ref{fig:lc}.  The maximum in the
0.2--1~keV band occurs at $\phi=0.94\pm0.05$, and at
$\phi=0.86\pm0.05$ in the 1--2~keV band.  While this indicates that
the maxima in the two bands are shifted by 0.08 in phase,
corresponding to a pulse shift of 29$^{\circ}$, the uncertainty on the
peak positions does not rule out the null hypothesis that there is
zero phase shift.  We note that this shift in phase is small compared
to the pulse width.  If the shift is real, this could be the origin of
the asymmetry in the 0.2--15~keV light curve.  

In an attempt to improve the sparse information we have from the sinusoidal
fit to the PN light curve in the 0.2--15~keV band, we have also modeled the
brightness variation with a symmetrical polar cap model \cite{crop01},
which includes gravitational bending effects.  The input parameters for the
model are the angular size of the caps, the angle between the magnetic and
rotation axis and that between the viewing direction and the rotation
axis.  The best fit to the data occurs for 20$^{\circ}$, 11$^{\circ}$ 
and 31$^{\circ}$ for the three angles respectively, with a
$\chi^{2}_{\nu}$ of 1.71.  The compactness parameter $M/2R$ was set to
a typical value of 3.  The $\chi^{2}_{\nu}$ for the polar cap model is
significantly better than for the sinusoidal fit with the same number
of free parameters, but falls short of that required because of the
symmetry of the model prediction. 

The absence of any sharp features in the light curve, and the absence
of any asymmetry in our model results in degeneracies between the fit
parameters and render the formal uncertainties on their values to be
large.  By exploring the parameter space, we find that a range of
viewing directions between 10 and 60$^{\circ}$ is permitted, with the
angle between magnetic and rotation axis constrained to between 20 and
35$^{\circ}$ over this range.  The change in viewing direction is in
fact accommodated by a change in cap size to produce the observed
pulsation, with the semi-angle subtended increasing from a negligible
size to 30$^{\circ}$ as the viewing angle increases from 10 to
60$^{\circ}$.  This analysis is therefore limited given the systematic
differences between the model and the data.  Nevertheless, the general
point that we note is that the method allows us to determine an upper
limit for the emitting region.  The pole cap size has to be
constrained to a full angle of $\sim$65$^{\circ}$, otherwise the
observed pulsation amplitude cannot be reproduced.

Another factor which could be responsible for the asymmetry is
aberration \cite{cordes78}.  This effect has been seen in radio
observations of pulsars \cite{gang01} and is due to corotation which
causes the radiation beams to bend toward the azimuthal direction (the
direction of rotation of the pulsar) so that emission is received
earlier than if there were no rotation.  The components emitted at
higher altitudes in the light cone are shifted to an earlier phase due
to this aberration, resulting in an asymmetry in the location of the
leading and trailing components with respect to the center of the
pulse profile.  For the same cone, the lower frequency radiation comes
from a higher altitude, and hence is observed first, than the higher
frequency radiation.  The effects of aberration increase with
altitude.  If we assume that the emission we observe from \pul\
originates from close to the surface of the neutron star i.e.\ a
heated polar cap, and not at a height which is near to that of the
light cylinder, the effects of aberration should be negligible.

\section{SPECTROSCOPY}
\label{sect:spec}

Data from MOS1 and PN were used to create spectra of \pul.  In order
to preserve any features in the spectra while providing sufficient
signal-to-noise, the spectra were re-binned to ensure a minimum of 15
counts in each energy bin.  We fit the combined MOS1 and PN spectra
with the X-ray spectral analysis package XSPEC (v11.1.0) using various
models, each with interstellar absorption ($N_{H}$) accounted for by XSPEC's
photoelectric absorption model. The spectra were fit between 0.2--3 keV.

We used three different methods to determine the value of $N_{H}$ to
be used in the model fits: (1) $N_{H}$ was allowed to run free, (2)
$N_{H}$ was fixed at the estimated fraction of galactic absorption in
the direction and at the distance of the pulsar with a value of
1.3$\times 10^{21}$ cm$^{-2}$ \footnote{This was achieved by determining the
galactic absorption in the direction of the pulsar \cite{lock90}, and
calculating the fraction of this column depth at the distance of the
pulsar.}, (3) $N_{H}$ was fixed at the value of 2.5$\times 10^{21}$
cm$^{-2}$ as measured from a star near to the pulsar in position and
distance. 

\pul\ is classed as a ``middle-aged'' pulsar and is thought to
perhaps have similar spectral properties to the {\it three Musketeers}
\cite{beck97,beck02b}.  The spectra of PSR~B0656+14 \cite{beck02b} and
B1055-52 \cite{pav02} are similar and can be well-fit with a model
which comprises of two blackbodies and a power law (PL), known as the
TS+TH+PL model.  The current interpretation is that the PL component
is due to some non-thermal magnetospheric emission, while the two
blackbodies represent a soft thermal component (TS) emerging from most
of the NS surface and a hard thermal component (TH) which originates
from heated polar caps.  Geminga is not bright enough to allow one to
discriminate between the different components of the temperature
distribution at the surface: its thermal spectrum is well-fit by a
single blackbody plus power law.   

In order to study the spectra of \pul, and to investigate the pulsar's
similarity to the {\it three Musketeers} we fit the PN and MOS1
spectra with a variety of models.  These were as follows, a single
blackbody, a single power law, a blackbody plus power law, a two
blackbody, and the TS+TH+PL model. 

We find that the multi-temperature models yield the same value for the
temperature for both components indicating that only one blackbody is 
required.  The large uncertainties on the power law index for the
single power law, blackbody plus power law and TS+TH+PL fit indicate
that the PL component is not required.  

\begin{figure}[h]
\epsscale{0.75}
\plotone{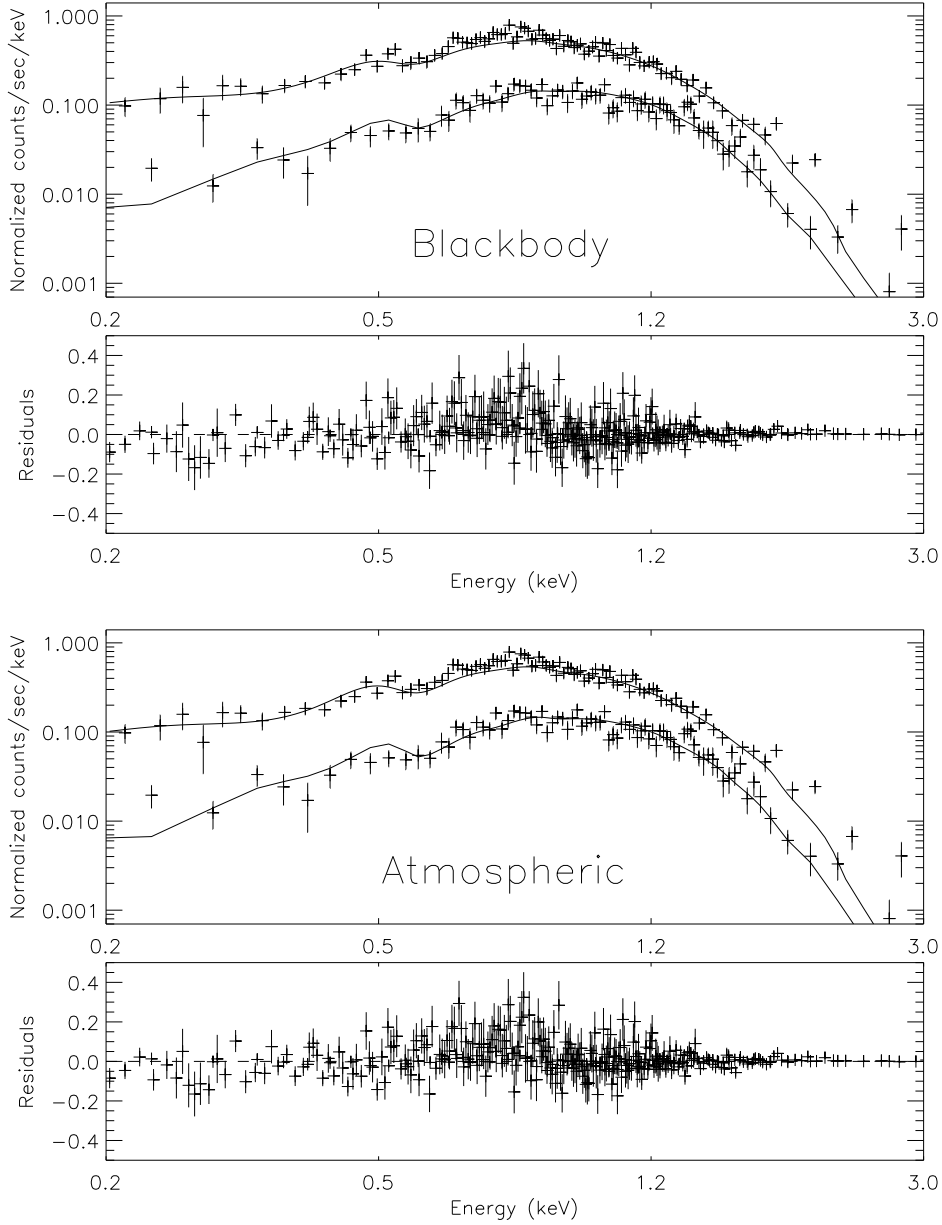}
\caption{PN and MOS1 spectra of \pul.  Top panel, solid line is a blackbody 
model with $kT$ = 0.183~keV and $N_{H}$ = 2.51$\times 10^{21}$
cm$^{-2}$, the $\chi^{2}_{\nu}$ for the fit is 1.08 for 321 d.o.f.
Bottom panel, solid line is a pure-H, non-magnetized atmospheric model
with $M_{NS}$ = 1.4 M$\sun$ and $R_{NS}$ = 10~km, 
$N_{H}$ = ($4.18^{+0.01}_{-0.06}$)$\times 10^{21}$ cm$^{-2}$ and
$T_{eff}$ = ($0.68^{+0.10}_{-0.01}$)$\times 10^{6}$ K, the
$\chi^{2}_{\nu}$ for the fit is 1.09 for 320 d.o.f.}\label{fig:spec}
\end{figure}

Our results indicate that the spectra can be well-fit with a single
blackbody (Fig.~\ref{fig:spec}, top panel).  Table 1 shows the results
for the fits with different $N_{H}$ values as described above.
Comparing the $\chi^{2}_{\nu}$ values for the three fits, and the
values of $N_{H}$, suggests that the estimate of the fraction
of galactic absorption in the direction of \pul\ is too low (fit (2)).

\begin{figure}[h]
\vbox to5in{\rule{0pt}{5in}}
\includegraphics{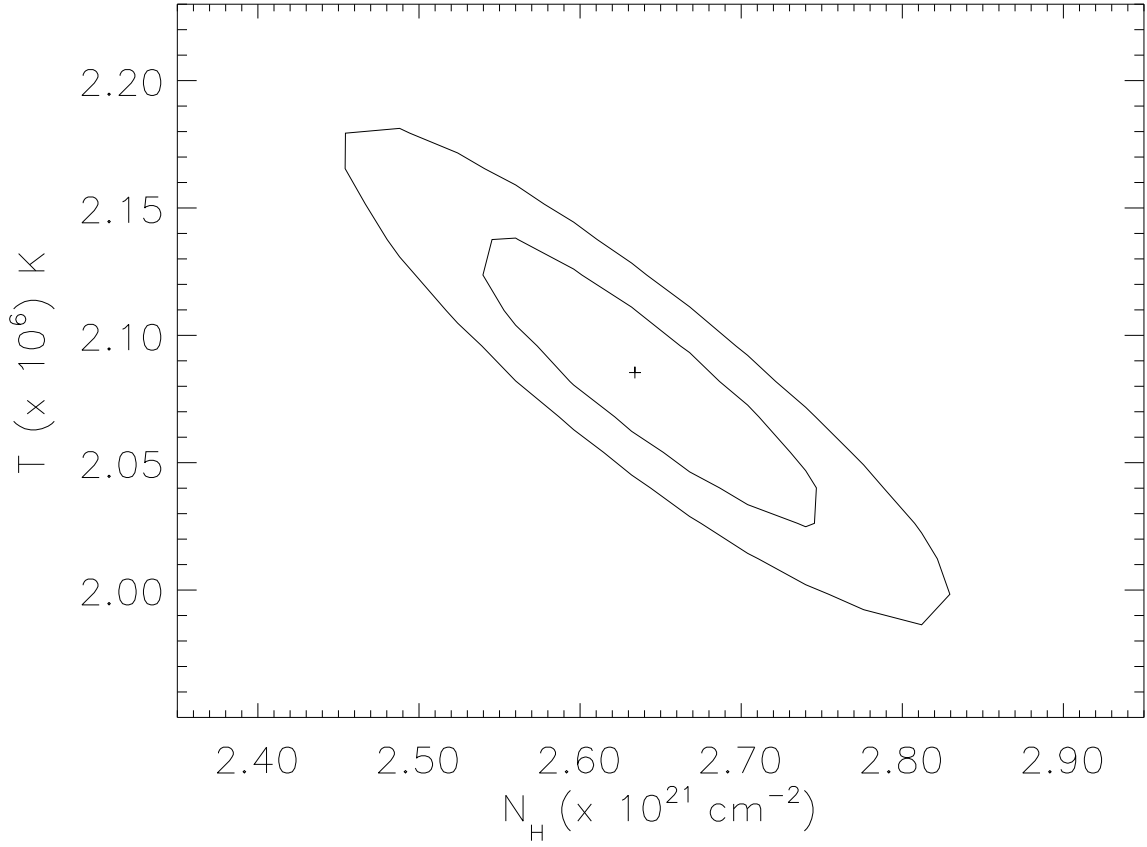}
\caption{Confidence contours in the temperature-absorbing column
plane.  Contours are for the 68\% and 90\% confidence levels.  The
best fit parameters from fit (1) in which $N_{H}$ was allowed to run
free is marked by a cross.}\label{fig:cont}
\end{figure}

From our model fit we find $N_{H}$ = ($2.63^{+0.02}_{-0.02}$)$\times
10^{21}$ cm$^{-2}$ and $T^{\infty} = (2.09^{+0.06}_{-0.07}) \times 10^6$ K
for fit (1) in which $N_{H}$ is allowed to run free.  The fit results in
$\chi^{2}_{\nu}$ = 1.08 for 320 d.o.f.  For fit (3) in which $N_{H}$
is fixed at the measured value for a star at a similar distance and
position as the pulsar of $N_{H}$ = 2.51$\times 10^{21}$ cm$^{-2}$ gives
$T^{\infty} = (2.12^{+0.04}_{-0.03}) \times 10^6$ K, with
$\chi^{2}_{\nu}$ = 1.08 for 321 d.o.f.  Employing the F-test we find a 
probability of 0.4 which indicates that freeing $N_{H}$ is not
warranted.  In Fig.~\ref{fig:cont} we show the error ellipses in the
temperature-absorbing column plane which confirm this result.

We find unabsorbed luminosities of 4.0$\times 10^{32}$ ergs s$^{-1}$
and 2.7$\times 10^{32}$ ergs s$^{-1}$ from XSPEC in the {\it ROSAT}
0.1--2.4 and 0.5--10~keV bands, and a blackbody luminosity
of 4.6$\times 10^{32}$ ergs s$^{-1}$ for fit (3).  The emitting radius
resulting from the single blackbody fit is $R^{\infty} = 1.68\pm0.05$~km.
This indicates that the emission originates from a region which is
smaller than the surface of the NS.

However, the radiation radius is representative of the stellar radius
only when the spectral properties of the emergent radiation are
consistently accounted for.  To investigate whether one can increase
the emitting radius so that it represents the whole of the neutron
star we have fit the data with a pure-H, non-magnetized atmospheric
model \cite{pav92,zav96} which is publicly available.  We note that
the observed modulation in the light curve argues against the emission
being generated by the full surface of the NS as implied by an
atmospheric model, however the size of the emitting region determined
from the model is dependent on the value of the normalization constant
and can therefore represent a smaller region.  Typically, these models
are found to require lower color temperatures and larger radii with
respect to a simple blackbody.  The results of the fits are given in
Table 2.  In the fits we let $N_{H}$ run free, but we fix the mass,
$M_{NS}$, and radius, $R_{NS}$, of the NS.  For $M_{NS}$ = 1.4 M$\sun$
and $R_{NS}$ = 10~km we find $N_{H}$ = ($4.18^{+0.01}_{-0.06}$)$\times
10^{21}$ cm$^{-2}$ and $T_{eff}$ = ($0.68^{+0.10}_{-0.01}$)$\times
10^{6}$ K, with $\chi^{2}_{\nu}$ = 1.09 for 320 d.o.f
(Fig.~\ref{fig:spec}, bottom panel).  For $M_{NS}$ = 1.4 M$\sun$ and
$R_{NS}$ = 18~km we find $N_{H}$ = ($4.17^{+0.01}_{-0.05}$)$\times
10^{21}$ cm$^{-2}$ and $T_{eff}$ = ($0.57^{+0.06}_{-0.01}$)$\times
10^{6}$ K, with $\chi^{2}_{\nu}$ = 1.10 for 320 d.o.f.  

\begin{figure}[h]
\vbox to6in{\rule{0pt}{6in}}
\includegraphics{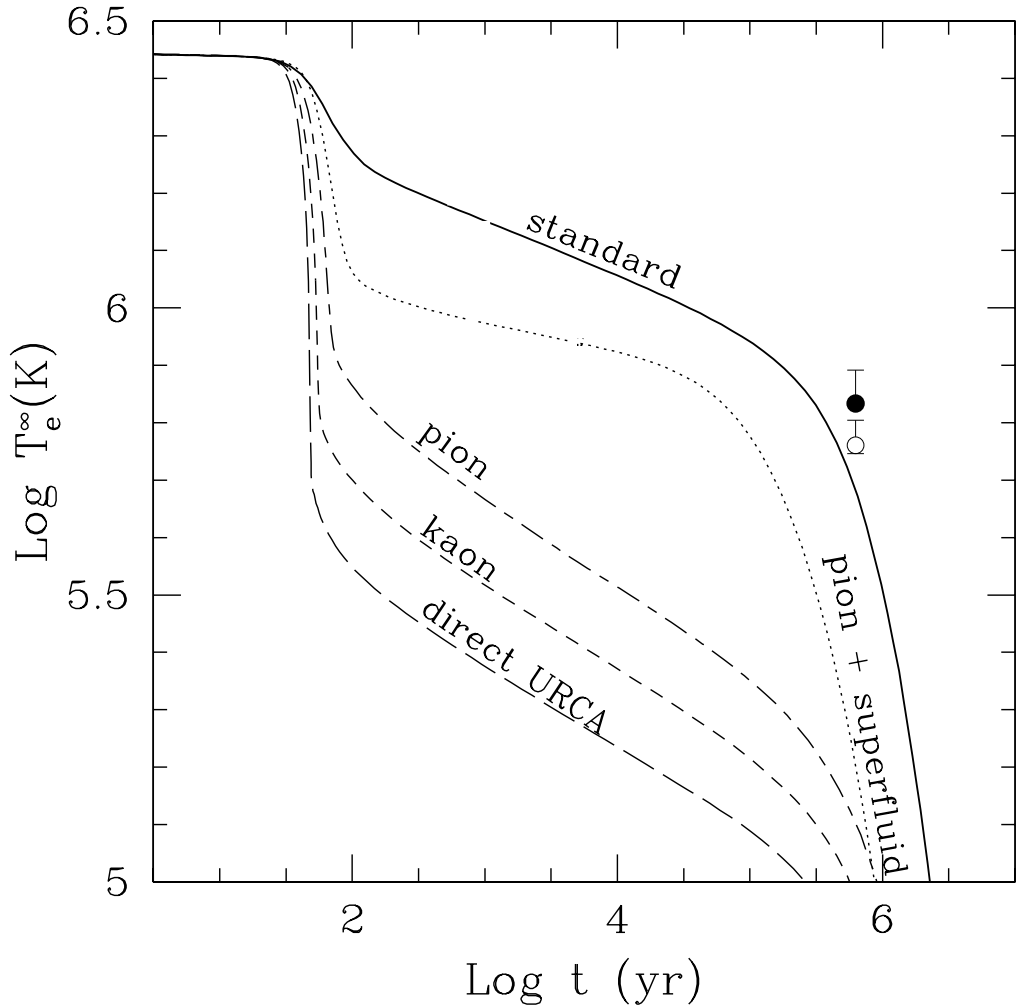}
\caption{Standard cooling curves (see e.g.~Tsuruta 1998; Page 1998).
Original figure from Slane et al.\ (2002).  The filled circle
corresponds to the temperature found from the atmospheric fit with R =
10~km, and the open circle corresponds to the temperature found
from the atmospheric fit with R = 18~km.  If error bar is not evident
it is encompassed in the data point.}\label{fig:slane} 
\end{figure}

The atmospheric model gives lower temperatures than the blackbody
model which are more in agreement with the predicted values from
standard cooling curves (see e.g.~Tsuruta 1998; Page 1998; Slane,
Helfand \& Murray 2002; see also Section \ref{sect:disc}).  The
standard cooling model is based on neutrino emission via the modified
URCA process, {\it n-n} and {\it p-p} neutrino bremsstrahlung, crust
neutrino bremsstrahlung, and plasmon neutrino processes.  Assumptions
about the form of the nucleon-nucleon force and the many-body
technique used to calculate the equation of state must be made in the
cooling models.  The standard curve considered here (see
Fig.~\ref{fig:slane}, solid line) uses an equation of state of
moderate stiffness.  However, the distance to the pulsar derived from
the atmospheric model fits are too low.  We find that the fit with
$R_{NS}$ = 10~km gives a pulsar distance of 0.26~kpc, and 0.38~kpc
with $R_{NS}$ = 18~km (see Section 1).  While this suggests that the 
atmospheric model can be ruled out, we note that the model we have
used is non-magnetized, and with $B\sim7\times 10^{11}$ G for \pul\ the 
applicability of this model is questionable.  Further studies with a
magnetic model must be performed to address this. 

We note that our spectral fits have been obtained from data which
are averaged over the spin pulse.  However, the pulsation can in principle
be caused by a local variation in the emission properties of the NS surface,
or by some absorption in a frame rotating with the NS.  These effects may
have implications for any spectral fitting to a mean spectrum.

\section{DISCUSSION}
\label{sect:disc}

\subsection{Results from the Spectral Analysis}

We have analyzed spectra of \pul\ from the MOS1 and PN instruments on
\xmm.  We find that the spectra are well described by a single
blackbody with $T^{\infty} = 2.12\times 10^6$ K and $N_{H}$ =
2.5$\times 10^{21}$ cm$^{-2}$.  The temperature inferred from the
model fit results in an emitting radius of $R^{\infty}=1.68$~km.  It is
difficult to reconcile this radius with that of the whole neutron
star.  The higher temperature implied by the blackbody fit could
indicate that there is a heating mechanism occurring at the poles
i.e.~magnetospheric processes, inverse Compton scattered photons above
the polar cap producing electron-positron pairs resulting in surface
heating by the returning positrons (see Greiveldinger et al.\ 1996;
Harding \& Muslimov 2002). 

An inferred temperature from a blackbody fit which results in a radius
which is smaller than the surface of the neutron star has been noticed
in several other sources \cite{beck02b}.  In general, fits with
un-magnetized, pure-H atmospheric cooling models give a temperature
lower by a factor 1.5--3, but often they also give radii that are too
large (or, equivalently, distances which are too small).  Therefore,
caution must be taken in whether they are directly applicable.  The
same is true in the assumption that the thermal emission is due to
accretion from the interstellar gas, since pure-H cooling and
accretion spectra are almost indistinguishable in the X-rays
\cite{tre02}.  If the atmosphere is dominated by heavy-elements, the
fit values from the model spectra are close to those from a blackbody,
but in this case one also expects absorption lines and edges
\cite{beck02b}.  

In the case of \pul, we have found that a pure-H, un-magnetized atmospheric
model also provides an excellent fit to the spectral data.  The resulting
temperature is a factor $\sim 3$ lower than that derived from the
blackbody fit.  The value is therefore high, but certainly in better
agreement with that predicted by the cooling models (see
Fig.~\ref{fig:slane}).  However, these fits imply either very small
distances to the pulsar ($\sim 200-400$~pc) or unacceptably large NS
radii, so the applicability of these spectral models to \pul\ is
questionable.  A similar conclusion has been reached by Pavlov et al.\
(2002) in the case of PSR~0656+14, based on \chandra\ data.  We note
that in the case of \pul\ fitting with a magnetized must be performed
to be able to rule out conclusively an atmospheric model.

\pul\ has often been claimed to be potentially similar to other
middle-aged sources, PSR~B0656+14, B1055-52 and Geminga (the {\it three
Musketeers}, see Table \ref{tab3}).  As noted in
Section~\ref{sect:spec} the power law component is detected in the
spectra of the {\it three Musketeers}.  Results from {\it ROSAT}
indicate that all of the three objects exhibit a pulse shift of 
$\sim 100^\circ$ at $E\sim 0.4-0.6$~keV, and an increase in the pulse
fraction (confirming the different nature of the thermal and
non-thermal emission).  However, recent results for B1055-52 deduced
from \xmm\ observations indicate that the pulse phase energy
dependence is not as strong as suggested by {\it ROSAT}.  The surface
temperatures of the {\it three Musketeers} (as taken by the blackbody
fit) are in the range (0.3--1.2)$\times 10^{6}$ K which is typically
too high to be reconciled with that predicted by standard cooling
curves \cite{beck02b}.  We note that the temperature determined for
\pul\ from the blackbody fit is even higher.  The inferred radii for the
{\it three Musketeers} are compatible with the equation of state of a
neutron star, or smaller.  However, as discussed above this issue is
extremely delicate since the values of peak temperature, star radius
(and distance) are strongly dependent on the assumption about the
spectral model. 

In the light of this situation, \pul\ is an extremely interesting
object.  Its spectrum {\it only} shows a thermal component: the
non-magnetospheric power law has not been detected by \xmm.  This result
must be confirmed by further deep pointings: for instance, we note
that in the case of B1055-52 the {\it ROSAT} spectrum was well-fit by
either two blackbodies \cite{gre96} or a flatter power law \cite{wa98}.
Only very recently have \chandra\ observations of B1055-52 \cite{pav02}
revealed the spectral energy distribution in detail; the \chandra\
spectrum is well-fit by the TS+TH+PL model.  We note that B1055-52 is
at a distance of $\sim 0.7$~kpc and faint, \pul\ is also faint and at
a larger distance, hence both sources are difficult to observe.

On the other hand, if confirmed, the absence, or very low level, of a
hard tail in the spectrum of \pul\ is peculiar.  While \pul\ and
B1055-52 have similar ages, \pul\ may represent a transition object
between those discussed above (PSR~B0656+14, B1055-52 and Geminga) and
older ($>10^6$ yrs) sources (e.g. B0950+08, B1929+10, J0437-4715),
based on its spectral properties.  The latter have a NS surface
(excluding the cap) which is much cooler and therefore not detectable
in the X-ray range.  However, the polar caps may be heated by
accretion or bombardment of particles thus producing the observed
X-rays.  In this case, the soft X-ray spectrum may be dominated by the
TH component discussed above.  This interpretation may also help to
reconcile the temperature inferred from the spectral fit of \pul\
which is higher than the predicted surface temperature of a NS at the
age of \pul\ (and makes the comparison with the cooling curves unfounded).

If the temperature from our blackbody fit to the spectrum of \pul\ is
taken as the TH component, we find an upper limit, from the 90\%
uncertainty level, of 0.56$\times 10^{6}$ K for the surface
temperature (TS) of the neutron star.  We find that the 2$\sigma$
upper limit for the unabsorbed flux in the 3.0--10.0 keV band for a
power law with a photon index of 2.0 is 5.9$\times 10^{-14}$ erg
cm$^{-2}$ s$^{-1}$.

\subsection{The Pulse Profile}

The \xmm\ observation of \pul\ provides the first evidence for pulsed
X-ray emission at a period which is consistent, within errors, to the
predicted radio pulse period.  The pulse profile is broad and slightly
asymmetric, with a slower rise and more rapid decline.

Pulsed X-rays from \pul\ could be produced via one of two mechanisms.
The first is due to relativistic particles produced either in regions above
the polar caps of the neutron star or in outer gaps.  These particles
are accelerated by the magnetic field which is aligned with the
electric field and/or the drop in potential generated by the Deutsch
wave.  Non-thermal X-ray emission results from the effects of
curvature, synchrotron and inverse Compton radiation on the
accelerated particles.  The other mechanism is thermal emission which
may be produced at the stellar surface due to the initial cooling of
the neutron star, or reheating of the polar caps by the bombardment of
back-flowing particles and internal friction.

The particular X-ray emission mechanism can sometimes be determined
from the pulse shape.  High amplitude, sharp, narrow pulses
can only be produced by a highly beamed, and therefore relativistic
population of electrons, while low amplitude, quasi-sinusoidal pulses
can be produced by either method.  Since the latter characteristics are
those seen in the PN light curve of \pul, in principle we cannot discriminate
between the two mechanisms.  However, the absence of a power law tail in
the spectrum, argues toward a thermal origin.

The observed pulsation may also be due to absorption in the rotating
magnetosphere, instead of being an intrinsic characteristic of the pulsar
emission (see Cropper et al. 2002 for a discussion of a similar
effect in the case of RX~J0720--3125).  However, the thermal emission
of \pul\ is hotter than that of RX~J0720--3125, and this implies a
higher ionization parameter (by a factor of $\sim 20$).  Moreover
\pul\ has a faster rotation period (by a factor of $>$ 50) which
limits the maximum extent of the light cylinder and therefore the
volume of absorbing material.  The high densities of largely neutral
material required, together with the rather weak magnetic field in
this system to confine it, argue against absorption as the origin of
the pulsations.

The simplest explanation for the flux variation is therefore that it
is due to hot, rotating polar caps.  The asymmetry in the pulse shape in
Fig.~\ref{fig:lc} (top panel) cannot be produced by any symmetrical
structure on the NS, nor by any beaming effects, as long as these
operate symmetrically with respect to the magnetic or rotation axis.
On the other hand, if the emission originates in a hot spot (or cap),
the observed asymmetry can be qualitatively understood by requiring
that the trailing edge of the cap is hotter and brighter than the
leading edge.  The DC component of the pulse profile implies that
there is never a time when the emission is ``off'' (see
Fig.~\ref{fig:lc}).  If the polar cap model is correct this suggests
that part of the cap is always visible.  

\section{CONCLUSIONS}
\label{sect:conc}

We find that the spectra of \pul\ can be well-fit with a blackbody.  
The value for the emitting radius indicates that the emission is from
a region which is smaller than the surface of the NS, i.e. a hot spot.
We investigated whether a pure-H atmospheric model would provide model 
temperatures to be in line with that expected from the cooling curve.
We find that the model fits the spectra well, but the distance to the
pulsar determined from the fit is too small, unless we use
unrealistically large radii for the NS.  We do not find any evidence
for a non-thermal component in the spectra of \pul.  However, we note
that the power law component was detected in B10522-52 only after
deeper observations.  This may also be true for \pul. 

We have detected for the first time pulsed X-ray emission from \pul,
the period of which is consistent, within errors, with the predicted
radio period.  Whether the mechanism for X-ray emission is via
non-thermal magnetospheric synchrotron or thermal emission from the hot
surface is not clear, as the broad asymmetric pulse shape found
for \pul\ does not favor either model.  However, the absence of a power
law component may indicate that the emission is thermal in origin.  

We find an asymmetric pulse profile which may be due to hot, rotating polar
caps, with a temperature structure which is hotter at the trailing
edge.  If the polar cap model is correct, the small radius we
determine for the emission region is expected.

\section{ACKNOWLEDGMENTS}

This work is based on observations obtained with \xmm, an ESA science
mission with instruments and contributions directly funded by ESA
Member States and NASA.  The authors acknowledge support from the
Institute of Geophysics and Planetary Physics (IGPP) program at LANL
and NASA grants S-13776-G and NAG5-7714.  K.~McGowan acknowledges
S.~Trudolyubov, and S.~Zane acknowledges R.~Turolla, for helpful
discussions.

\clearpage

\clearpage

\clearpage

\clearpage

\clearpage

\clearpage

\clearpage
\begin{deluxetable}{cccccc}
\tablewidth{29pc}
\tablecaption{Blackbody fits to \pul
\label{tab1}}
\tablehead{
\colhead{Fit} & \colhead{$N_{H}$}  & \colhead{$kT$} &
\colhead{$\chi^{2}_{\nu}$ [dof]} & \colhead{$T^{\infty}$} 
& \colhead{$R^{\infty}$} \\
\colhead{} & \colhead{$\times 10^{21}$ cm$^{-2}$}  & \colhead{keV} &
\colhead{} & \colhead{$\times 10^{6}$ K} & \colhead{km}}
\startdata
1 & $2.63^{+0.02}_{-0.02}$ & $0.180^{+0.005}_{-0.006}$ & 1.08 [320] &
$2.09^{+0.06}_{-0.07}$ & $1.79^{+0.13}_{-0.10}$ \\
2 & 1.30 (fixed) & $0.216^{+0.003}_{-0.003}$ & 1.42 [321] &
$2.50^{+0.04}_{-0.03}$ & $0.87^{+0.03}_{-0.02}$ \\
3 & 2.51 (fixed) & $0.183^{+0.003}_{-0.003}$ & 1.08 [321] &
$2.12^{+0.04}_{-0.03}$ & $1.68^{+0.05}_{-0.05}$ \\
\tablecomments{Fit 1: $N_{H}$ is allowed to run free, fit 2: $N_{H}$
fixed at the estimated fraction of galactic absorption in the
direction and at the distance of the pulsar, fit 3: $N_{H}$ fixed at
the measured value from a star near to the pulsar in position and distance.}
\enddata
\end{deluxetable}

\clearpage
\begin{deluxetable}{cccccc}
\tablewidth{30pc}
\tablecaption{Pure-H, non-magnetized atmospheric fits to \pul
\label{tab2}}
\tablehead{
\colhead{$N_{H}$} & \colhead{$M_{NS}$} & \colhead{$R_{NS}$}
& \colhead{$T_{eff}$} & \colhead{$\chi^{2}_{\nu}$ [dof]} &
\colhead{$D$} \\
\colhead{$\times 10^{21}$ cm$^{-2}$} & \colhead{M$\sun$} & \colhead{km}
& \colhead{$\times 10^{6}$ K}  & \colhead{} & \colhead{kpc}}
\startdata
$4.18^{+0.01}_{-0.06}$ & 1.4 (fixed) & 10 (fixed) & $0.68^{+0.10}_{-0.01}$
& 1.09 [320] & 0.26 \\
$4.17^{+0.01}_{-0.05}$ & 1.4 (fixed) & 18 (fixed) & $0.57^{+0.06}_{-0.01}$
& 1.10 [320] & 0.38 \\
\enddata
\end{deluxetable}

\clearpage
\begin{deluxetable}{lccccc}
\tablewidth{33pc}
\tablecaption{Pulsar Parameters for the {\it three Musketeers} and \pul\
\label{tab3}}
\tablehead{
\colhead{Pulsar} & \colhead{$P$}  & \colhead{$\dot P$}  &
\colhead{$\log \left (P/2 \dot P \right )$} & \colhead{$D$} &
\colhead{$\log{B}$} \\
\colhead{} & \colhead{ms}  & \colhead{$10^{-15}$ s/s} &
\colhead{yr} & \colhead{kpc} & Gauss \\
\colhead{} & \colhead{} & \colhead{} &
\colhead{} & \colhead{} & \colhead{}}
\startdata
B0633+17 (Geminga) & 237.09 &10.97 &5.53 & 0.16& 12.21 \\
B0656+14 & 384.87 & 55.03 & 5.05 & 0.67& 12.67 \\
B1055-52 & 197.10 & 5.83 &5.73 & 0.73& 12.03 \\
\pul & 143.15 & 3.66 &5.79 & 1.20 & 11.87 \\
\tablecomments{For all references we refer to Becker \& Tr\"umper (1997).
The magnetic field is computed from the spin-down formula.  The pulsar
distances are computed according to the model of Cordes \& Lazio (2002).}
\enddata
\end{deluxetable}

\end{document}